\documentclass[12pt,preprint]{aastex}

\begin{document} 
\title{Structural Parameters for Globular Clusters in NGC 5128. II: HST/ACS Imaging and New Clusters\footnote{
Based on observations made with the NASA/ESA Hubble Space Telescope, 
obtained at the Space Telescope Science Institute, which is 
operated by the Association of Universities for Research in Astronomy, Inc., 
under NASA contract NAS 5-26555. These observations are associated with program \#10260.
Support for program 10260 was provided in part by NASA through a grant from the Space 
Telescope Science Institute, which is operated by the Association of 
Universities for Research in Astronomy, Inc., under NASA contract NAS 5-26555.}
}

\author{William E. Harris} 
\affil{Department of Physics \& Astronomy, McMaster University, Hamilton L8S 4M1, Canada}
\email{harris@physics.mcmaster.ca} 

\author{Gretchen L.~H.~Harris}
\affil{Department of Physics \& Astronomy, University of Waterloo, Waterloo N2L 3G1, Canada}
\email{glharris@astro.uwaterloo.ca}

\author{Pauline Barmby}
\affil{Harvard-Smithsonian Center for Astrophysics, 60 Garden Street, Mail Stop 65, Cambridge, MA 02138}
\email{pbarmby@cfa.harvard.edu} 

\author{Dean E. McLaughlin}
\affil{University of Leicester, Department of Physics and Astronomy, University Road, Leicester LE1 7RH, UK}
\email{dean.mclaughlin@astro.le.ac.uk}

\author{Duncan A. Forbes}
\affil{Centre for Astrophysics \& Supercomputing, Swinburne University, Hawthorn, VIC 3122, Australia}
\email{dforbes@astro.swin.edu.au}

\shorttitle{Clusters in NGC 5128}
\shortauthors{Harris et al.}

\begin{abstract} 
We report the first results from an imaging program with the ACS camera
on HST designed to measure the structural characteristics of a wide range
of globular clusters in NGC 5128, the nearest giant elliptical galaxy.
From 12 ACS/WFC fields, we have measured a total of 62 previously known
globular clusters and have discovered 69 new high-probability cluster candidates not
found in any previous work.  We present magnitudes and color indices for all of
these, along with rough measurements of their effective diameters and ellipticities.
The luminosity distribution of this nearly-uncontaminated sample of
clusters matches well with the normal GCLF for
giant elliptical galaxies, and the cluster scale size and ellipticity
distributions are similar to those in the Milky Way system.  
The indication from this survey
is that many hundreds of individual clusters remain to
be found with carefully designed search techniques in the future.
A very rough estimate of the total cluster population from our data suggests
$N_{GC} \simeq 1500$ in NGC 5128, over all magnitudes and within a projected 
radius $R = 25'$ from the galaxy center.
\end{abstract}


\section{Introduction} 
\label{intro}

Interest in the structural ``fundamental plane'' (FP) for globular clusters 
\citep{dj95,mcl00,mcl05} has
grown rapidly over the past few years.   We can now 
explore the structure of such objects in great detail, and in several
types of host galaxies, over a range of almost 4 orders of magnitude in
cluster mass.  Furthermore, there is
growing evidence that the uppermost end of the globular cluster (GC) mass range 
($10^7 M_{\odot}$ and even beyond) is inhabited by a wider variety of objects than
was previously realized.  The FP region above $10^6 M_{\odot}$ is populated with objects
easily categorized as classic globular clusters, but also by structures such as the compact
nuclei of dwarf elliptical galaxies, ``extended luminous'' clusters \citep{huxor},
and the new class of Ultra-Compact Dwarfs \citep[e.g.][]{phil01, has05, jones06},
themselves perhaps a mixed collection of objects.  Some of these other types of
objects are proving hard to distinguish from normal GCs at large distances,
because their effective radii and $M/L$ ratios are only slightly larger.
It is also this mass range at which the downward extrapolation of the E-galaxy
scaling relation intersects the GC locus \citep{drink03, has05}.

The nearest galaxy containing large numbers of these most massive globular 
(or globular-like) clusters is NGC 5128, the giant elliptical at the center of 
the Centaurus group at a distance of just 4 Mpc.   With the help of high-resolution
imaging, we have the opportunity to define the FP of a globular cluster system 
accurately and homogeneously over its
entire mass range and within a single galaxy, as well as to study GCs within a giant
elliptical individually and in unequalled detail.   Defining as large as possible
a sample of individual GCs is also a prerequisite to many other kinds of 
investigations, including the kinematics and dynamics of the cluster system,
its metallicity distribution, and cluster ages, all of which are needed to build up
a complete understanding of its evolutionary history \citep[e.g.][]{hghh92,peng04,harris04,woodley05}.

HST-based imaging of GCs in NGC 5128 was carried out by \citet{harris98} for a single
outer-halo cluster, and by \citet{holland99} for a selection
of inner-halo clusters; these studies were both done with the WFPC2 camera.
Subsequent work by \citet{rej01} and \citet{gomez06} has shown that ground-based
imaging at $\sim 0\farcs5$ seeing conditions can be used to identify GCs 
with high probability and even to measure key structural parameters
(half-light diameters and central concentrations).  Identifying cluster candidates
in NGC 5128 this way is, in fact, far more efficient than other techniques such as
color indices and radial velocity surveys because of the high degree of field
contamination at the Galactic latitude of this galaxy 
\citep[see][for more extensive discussion]{harris04a}.
     
The superior resolution of the cameras on board HST is, however, the only
available way to measure these clusters at high enough resolution to establish
their core radii and even central surface brightnesses with some confidence.
In \citet{harris02}, hereafter referred to as Paper I, we used data 
from the WFPC2 and STIS cameras to study 27 individual clusters in NGC 5128.
Velocity dispersions and mass estimates have been obtained for 14 of the
very most luminous of these
by \citet{mar04}.  In this paper, we present new imaging data for 
a much larger sample of 131 GCs, 69 of which are newly discovered.

\section{Imaging Observations}
\label{observations}
\begin{figure} 
\figurenum{1} 
\plotone{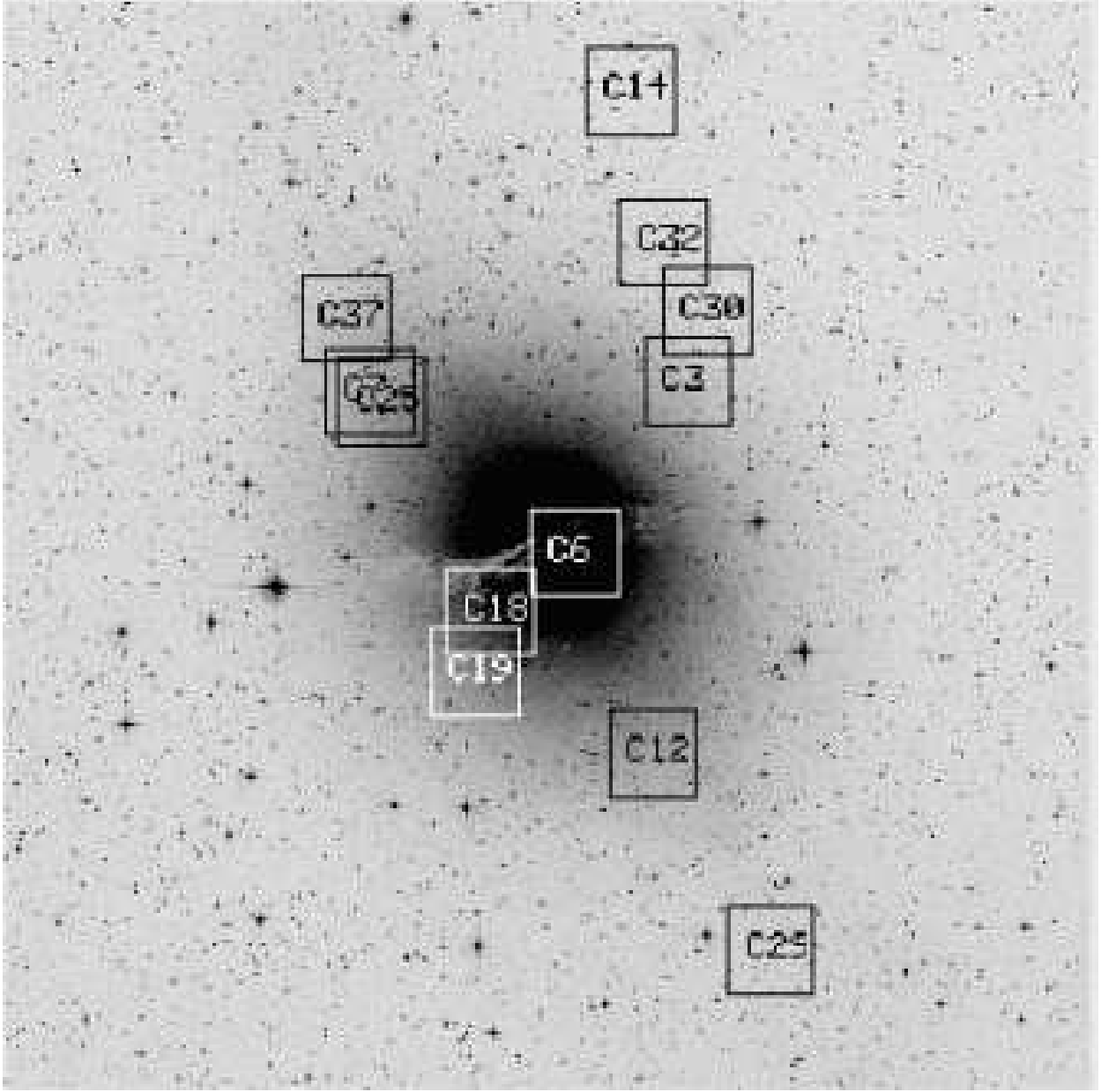} 
\caption{Location of our ACS target fields relative to
NGC 5128.  The field size shown is $42'$ across, and North is
upward and East to the left.  The size of
each box corresponds to the $3\farcm3$ size of the ACS/WFC field.
The boxes are meant only to mark the field center locations
relative to the galaxy center, but do not show their orientations,
which differ individually.}
\label{chart}
\end{figure}

Our HST program 10260 was 
targetted at imaging many of the most massive known GCs in both
M31 and NGC 5128.  The M31 material will be presented in a later paper
(Barmby et al., in preparation).  In NGC 5128, 12 fields were imaged, centered
on the clusters C3, C6, C7, C12, C14, C18, C19, C25, C29, C30, C32, and C37 
\citep{hghh92}.  We used the ACS camera in its Wide Field Channel
(field size $3\farcm3$ and scale $0\farcs05$ per pixel), and with a
total exposure time on each field of one full orbit of 2500 sec divided into three
equal sub-exposures, in the F606W (``Wide $V$'')
filter.  In Figure \ref{chart} we show the locations of these 12 target
fields around NGC 5128. Our intentions in employing these moderately long exposures with the
wide field were, first, to allow us to trace the cluster envelopes to 
significantly larger radii than we could do in Paper I; and second, to add to 
our sample by identifying new GCs in the fields.  Because the brightest 
red-giant stars in the halo of NGC 5128 (and in its clusters) are resolved
on these images, our chances of 
identifying clusters versus contaminating objects
such as background galaxies with roughly similar sizes are much improved
over even the best ground-based imaging.

\section{Identification of Clusters and Cluster Candidates}
\label{Identifications}

For all the data analysis we used the combined, multidrizzled images taken
from the HST Archive.  These images are corrected in scale and flux
normalizations for the ACS geometric distortion.  
Three separate passes of visual inspection combined
with IRAF/{\sl imexamine} were made
on each frame, to search for clearly nonstellar objects that were also noticeably
resolved into stars themselves and to obtain their profile widths
and approximate shapes.  Objects with FWHMs smaller than $\sim 3$ px (i.e. slightly
larger than the stellar profile width of $\simeq 2.4$ px), 
{\sl and} that were not resolved into
stars, were rejected.  We also rejected objects clearly redder than
the color range enclosing normal GCs (see below); these red objects were
not resolved into stars, and most had structural features and asymmetries marking
them as probable background galaxies.  
Finally, the complete list of previously known GCs
from the literature was checked to make sure that our final list included all 
62 that turned out to fall within our field boundaries.  Our measurement procedure had
already independently rediscovered all but half a dozen of these,
indicating that our detection completeness rate for finding real clusters
was above 90\%.  The handful that were missed have very compact structures with
FWHM values barely larger than the stellar point spread function; in principle,
automated and more quantitative profile measurements for all the objects on the
frames might have recovered these, but at the risk of increasing the contamination
of the sample by foreground stars.

Over all 12 fields, we found a total of 62 previously known GCs, along with 69
new objects that are not in any previous list and that we regard
as highly probable GCs.  This set of objects greatly increases
the HST-based imaging sample of clusters for which accurate structural
parameters can be obtained, and provides a strong basis for
defining the GC fundamental plane in NGC 5128.  In Paper III (McLaughlin et al.,
in preparation), we will discuss those results in full.  Here we present an  
overview of the sample.

Images for all the individual clusters in our ACS frames are shown in Figures 2a,b,c.    
These clearly show the wide range of luminosities they have, as well as a
noticeable range in ellipticity and effective radius.  
The single cluster whose integrated colors might be affected adversely by
differential reddening from the central dust lane in NGC 5128 is C150 (see
again Fig.2b; all the others are in clear regions of the halo (or, in the case
of field C6, the bulge).  Any candidates that were {\sl very} heavily affected
by differential reddening would, in fact, have been pushed beyond our adopted
color limits and rejected from the sample.  For a more extensive discussion
on the reddening effects within the bulge region, see \citet{hghh92}.

In Table 1 we list the previously known GCs falling within our fields. 
The successive columns of the Table include the object
coordinates (J2000); position $(R, \theta)$ relative to the center of NGC 5128 (where
radius $R$ is in arcminutes and
azimuthal angle $\theta$ is measured East of North); photometric indices in the
Washington system, taken from the lists of \citet{hghh92} and \citet{harris04}
(hereafter referred to as HHG04);
and the ellipticity $e$ and Moffat-profile FWHM (in pixels) as obtained from {\sl imexamine}.
Conveniently, at the 3.9-Mpc distance of NGC 5128, one pixel of $0\farcs05$ corresponds
to a linear scale of very close to 1 parsec.
For the cluster identification numbers, the `C' and `G' clusters are from \citet{hghh92}; PFF numbers from
\cite{peng04}; WHH numbers from \citet{woodley05}; f1 and f2 numbers from 
\citet{rej01}; AAT numbers from \citet{beasley06}; and K numbers from \citet{kraft01}.

In Table 2, we give the same information for the 69 newly discovered cluster candidates.
These are labelled continuing with the C-numbers from \cite{hghh92} and Paper I, 
starting at C111.  The last two columns list the ACS field on which each one
appeared, and any notes of peculiarities.  

During the process of inspecting the previously cataloged clusters on our images,
we found two that were clearly foreground stars, not clusters (they have stellar
FWHMs, are unresolved, and show faint diffraction
spikes):  these were PFF010 and AAT114993.  Interestingly, these two have measured
radial velocities of 344 and 352 km s$^{-1}$, just in the velocity range where
NGC 5128 clusters can overlap with high-velocity Milky Way halo stars 
\citep{peng04, woodley05}.  Their magnitudes and colors fall within
the normal range of the NGC 5128 GCs ($T_1 = 19.24, (M-T_1) = 0.67$ for PFF010;
$T_1 = 19.83, (M-T_1) = 0.98$ for AAT114993).

A few of our ACS target fields overlapped each other, such that 16 of our 
GCs fell on more than one frame.  These allowed useful consistency checks of the measured
profile widths and ellipticities.  We found that both $e$ and the FWHM agreed
internally between frames extremely well, with rms scatter of just $\pm 0.003$ in $e$
and $\pm 0.09$ px in FWHM.  

Some of the new clusters (Table 2) do not appear in the HHG04 database, in most
cases because they lie relatively close to the galaxy center in the bulge region
not covered in their work.  These are the ones without $(M-T_1)$ or $(C-T_1)$ color
indices in Table 2.  To estimate approximate magnitudes for them, we used
the aperture magnitudes from {\sl imexamine} and correlated them with the $T_1$
values for clusters in common with HHG04.  This mean shift was then applied to
$m(imexam)$ to get a rough value for $T_1$; however, these should not be considered as
accurate to better than $\pm 0.3$ mag.

The astrometric zeropoints for our listed 
coordinates are not those in the raw ACS image headers,
which can be wrong by an arcsecond or more.  We applied small offsets to the world
coordinate systems of these images so that the positions of stars in the fields
matched those in the 2MASS point source catalog.  These revised coordinates were
further checked against the database of \citet{harris04a}, which gives coordinates 
of all objects in a one-degree field centered on NGC 5128 and tied to the 
USNO UCAC1 system with a claimed accuracy of
$\pm 0\farcs2$ \citep[see][for extensive discussion]{harris04a}.
Our object coordinates listed in Table 1 and 2
agree with that system to within a median $\Delta r = 0\farcs18$,
quite consistent with the expected accuracy.

As a final note on the coordinates, we found through these
comparisons that the $(\alpha, \delta)$ values
for the clusters in our Paper I -- which were determined from the
STIS image headers -- 
were systematically wrong by more than an arcsecond, with most
of the error coming in declination.  The coordinates in
Paper I need to be corrected by $\Delta \alpha = (+0.040 \pm 0.020)$ seconds of time
and $\Delta \delta = (+1.4 \pm 0.20)$ seconds of arc.  We have calculated
these shifts by rematching the 27 separate clusters in Paper I with their
coordinates in the HHG04 database.  These ($\Delta \alpha, \Delta \delta$)
values are confirmed by the five clusters in the present ACS imaging
study that overlap with the ones in Paper 1.  For completeness, in Table 3
we provide the revised coordinates and the Washington photometric indices
for the clusters we observed in Paper I.

\section{Discussion of Results}

In Figure \ref{fig2}, we show the distributions in magnitude ($T_1$) and 
colors ($M-T_1)$ and $(C-T_1)$ for all 149 of the objects that we have
imaged under high resolution with the HST cameras (Tables 1 to 3).
The classically bimodal color distribution of GCs, already well known in
NGC 5128 \citep[e.g.][]{hghh92,rej01,woodley05} as in other giant galaxies, shows up in the 
$(C-T_1)$ graph particularly (see Woodley et al.~2005 for histograms
with the same color indices and a larger database).  It is apparent from Fig.~\ref{fig2} that this
method of identifying clusters, first by image morphology and second by color, 
does not discriminate
against the fainter clusters as much as radial velocity studies have done.
In other words, this sample is more uniformly distributed in magnitude than those
identified with velocities (to date, the radial velocity data for the NGC 5128
clusters are all from 4m-class telescopes, and these samples have effective
faint-end limits near $V \simeq 21$; see Peng et al.~2004a and Woodley et al.~2005).  
In particular, the newly discovered cluster candidates
are mostly fainter than $T_1 \simeq 20$, with many of them fainter than the
classic GCLF turnover point.  This work suggests that there are likely to be
many hundreds of globular
clusters remaining to be found in NGC 5128, and that high-resolution imaging
is an effective tool for isolating candidates.  A very small number of objects
are noticeably bluer than the normal old-halo clusters, possibly indicating 
younger ages for these.

In Figure \ref{fig3}, we show the distribution of FWHM values from Tables
1 and 2.  Many of our clusters and candidates have effective diameters in
the range of $ 3 - 4$ pc or less and would be quite hard to resolve
under ground-based seeing conditions.  The fact that the histogram
keeps rising almost up to the PSF diameter of $\simeq 2.4$ px (dashed line)
indicates that we may still be missing some clusters with extremely small
effective diameters.  Such objects will be quite hard to find except perhaps
through radial velocity surveys where all objects in a given region
are targetted regardless of morphology.   Although
the FWHM values should be taken only as a rough estimate of the effective
diameters, the distribution already resembles that for the Milky Way
globular clusters shown in the lower panel of the Figure, where the
majority of the globular clusters have half-light diameters in the range
$2 - 6$ pc.  It is worth noting as well
that the histogram of cluster diameters has a long, extended tail
to larger radii, and that we have found a significant number of such
objects in our ACS survey; these could easily be picked out from ground-based
imaging in sub-arcsecond seeing conditions and should be more carefully
looked for in a more extended survey, particularly in the outer halo.

In Figure \ref{fig4}, the FWHM estimates
are plotted against galactocentric radius.  
In general, the clusters with smaller diameters tend to be found
preferentially closer to the galactic center, although the trend
is not strong and many
rather extended clusters are visible at all radii.
Similarly, there is a weak trend for the FWHM to be larger for the
bluer (more metal-poor) clusters.
Our data follow the same general trends of GC size versus galactocentric radius
and color as in the comprehensive analysis of the Virgo galaxies by
\citet{jor05} (see also Larsen et al. 2001 for an
earlier study of GCs in elliptical galaxies that found similar observational trends).
In our Paper III, we will present more rigorous effective radii for our cluster
sample based on full model fitting, where it will be possible to study these
correlations better than in our current (preliminary) dataset.

The great majority of the clusters in our ACS sample have low ellipticities,
roughly in the range of familiar globular clusters.
In Paper I we suggested from our much smaller sample of clusters imaged
with STIS and WFPC2 that NGC 5128 had proportionally more high$-e$
clusters than does the Milky Way.  The larger sample we have from our present
study is shown in Figure \ref{fig5}, including 120 clusters brighter than
$T_1 = 21$.  For comparison, 100 Milky Way clusters with known ellipticities
\citep{harris96} are shown as well.  These two samples are now very similar
in both size and distribution by ellipticity, and a two-sample test shows
no statistically significant difference between them.  Though this comparison
must be viewed with some caution because the $e-$values were measured in 
different ways for the two samples (see Paper I for additional discussion), 
we conclude that the GCs appear quite similar in shape in these two galaxies.
The few objects in Tables 1 and 2 that have $e > 0.3$
are all cases where the cluster is moderately faint and noticeably crowded by
neighboring, relatively bright stars.
Rejecting these few, we find a mean $\langle e \rangle = 0.08$ for the
clusters in our total sample.  The dropoff of the NGC 5128 histogram for $e < 0.04$
may not be significant, since
the average $e-$value for {\sl imexamine} measurements of 
{\sl stars} on our images is $\langle e \rangle \simeq 0.04$.
In Paper III, we will present the results of more rigorously derived
structural parameters, effective radii, and model profile fits.

A point of obvious interest in our new dataset is the 
globular cluster luminosity function
(GCLF).  As noted above, the ACS images reach deep enough that
clusters of almost all luminosities are about equally easy to find by the combination
of resolution into stars and profile measurement, with the exception of very
faint or extended ones in crowded starfields.  For this reason, it should be
possible to construct a relatively clean GCLF that is not 
strongly biased by luminosity.  \citet{rej01} used a similar approach with 
her ground-based imaging data, but our HST/ACS list is larger and more nearly
contamination-free.  HHG04 made an attempt
to construct a GCLF for the entire halo of the galaxy
on a statistical basis from their wide-field survey data,
but their results were severely compromised by the heavy field contamination.
By contrast, the only obvious bias that should exist in our sample is that the original frames
were deliberately targetted at individual bright clusters.  Other than these single
bright clusters, the objects found on the remainder of the ACS images constitute
a nearly unbiased sample from several different places in the halo, and should give
us a reasonable estimate of the GCLF over nearly its full range. 

In Figure \ref{fig6}, we show the results of this exercise.  If we combine all
the objects in Tables 1, 2, and 3, and then eliminate the 25 bright clusters that
were the deliberately chosen centers of each ACS and STIS image, we are left with
124 objects defining a nearly unbiased, nearly uncontaminated sample of GCs.
These are plotted in half-magnitude bins in $T_1$ as the solid dots in
the Figure.  Putting back the 25 bright target-center clusters gives the
open circles in the Figure, clearly showing the bright-end bias 
that they generate.  For comparison,
we show a ``standard'' GCLF curve for giant elliptical galaxies superimposed
on the data.  This is a Gaussian curve with standard deviation $\sigma = 1.3$ mag, and
a peak at $T_1 = 20.35$ corresponding to a turnover luminosity $M_V^{to} \simeq -7.4$
\citep{har01}, shown by the horizontal line in Fig.~\ref{fig2}.  The 
match of the standard curve to the data indicates that
a small amount of incompleteness on the fainter half of the distribution may still exist
in the sample. But in general, it suggests to us that the GCLF for NGC 5128
is at least roughly normal, falling easily into the same pattern 
seen in other galaxies.\footnote{The slight asymmetry of the plotted points relative
to the Gaussian curve (a) is not statistically significant particularly in view
of the possible faint-end incompleteness, and (b) in any case resembles what is
seen in the Milky Way and M31; see Harris (2001).  Since the shape of the bright half
of the GCLF is determined largely by the initial mass distribution of the clusters and
the faint half by both the initial numbers and
later dynamical evolution, there is no physical reason to expect
the GCLF to show exact symmetry when plotted in this form.}

Although the GCLF shown here is still constructed from a rather small sample, it is worth
noting that the GCLF for the Milky Way -- which for decades has acted as the baseline 
comparison system for all other galaxies -- is based on a sample of just the
same size.  In NGC 5128, however, there is obvious promise for building up a much larger
census of objects that is virtually free of contamination (as is also true for 
M31, though the total GC population there is likely to be only one-third that in NGC 5128).

\section{The Total Cluster Population in NGC 5128}

The total size of the globular cluster system in NGC 5128 has been extraordinarily
hard to gauge.  More than 20 years after
the first quantitative wide-field survey of the system \citep{harris84}, our estimates
of the total GC population $N_{GC}$ may still not be reliable
to any better than 50\% or so.
As is more thoroughly discussed in \citet{harris04a},
the essence of the problem is that the
large angular size of the NGC 5128 halo on the sky severely dilutes
the globular cluster population against the very
heavy field contamination by both foreground stars and faint background galaxies;
of all the objects in the $\sim 1^o$ field around the galaxy that are in the
same magnitude and color range occupied by the GCs, no more than about 2\%
are actually the clusters we want to find.
The statistical profile of the GC population derived from the wide-field 
CCD photometric survey of HHG04 yielded a total $N_{GC} = (980 \pm 120)$ clusters
over all radii, an estimate that can probably be viewed as a
firm {\sl lower} limit to the population since it adopted a 
conservative background-count level.
A fairly generous {\sl upper} limit \citep{harris84} is probably near 2000.  

The 12 ACS fields in our new survey give us another way, even if very rough,
to estimate the GC system profile and total population.  Although the total area
covered by these fields is quite small compared with the whole extent of the halo
(only 6\% of the whole area within $R \simeq 25'$; see below),
they are located at a wide range of radii and so can be used to gauge the radial
falloff of the cluster population.  Most importantly, the cluster sample they 
provide is a homogeneous and relatively ``pure'' one more nearly 
free of contamination than any previous material.  From the GCLF data
(see Fig.~\ref{fig6}), we can plausibly expect that we have found all
of the GCs in our frames down to at least a magnitude fainter
than the GCLF turnover level {\sl except} in the inner bulge regions where 
severe crowding and projections on the dust lanes reduce the discovery fraction.
Then if we plot up the number of detected GCs per frame against the projected
galactocentric distance of the frame, we should 
expect to see a radial falloff, albeit with much random scatter because
of small-number statistics.

A plot of this type is shown in Figure \ref{fig7}.  As expected, the innermost
field (C6 at $R = 1\farcm9$) appears to be heavily affected by incompleteness,
falling well below the trend set by the other 11 fields.  A least-squares
fit of a simple power law $n \sim R^{-\alpha}$ to these 11 points 
gives $n \simeq 550 R^{-1.8}$, shown by the solid line in Fig.~\ref{fig7}.
It is obvious from the Figure that these data are extremely sketchy, and 
the power-law slope $\alpha = 1.8$ is no better determined than
$\pm 0.3$; nevertheless, it is quite consistent with the value $\alpha = 2.0 \pm 0.2$
derived by HHG04, as well as with $\alpha = 1.6 \pm 0.3$ expected from
the known correlation of $\alpha$ with host galaxy luminosity (see HHG04).

Using the fact that each ACS/WFC frame has an area $\simeq 11.3$ arcmin$^2$,
and integrating the derived profile over the radial range $R = 3' - 25'$,
we then estimate a total of 1020 clusters over that range and brighter
than our adopted limit $T_1 \simeq 21.4$.  (Here we assume, more or less arbitrarily,
that the 10 clusters in our data fainter than that limit will compensate
for the slight incompleteness that we suspect may be present brighter 
than that level, as noted above.  Within the accuracy
of our calculation, this correction is negligible.)  Inside $R \simeq 3'$ the density 
profile $\sigma(R)$ of the GC system flattens off (HHG04), so to estimate
the number of clusters in the core we take a constant
$\sigma(3') = 6.7$ arcmin$^{-2}$ over that area, giving a further 190
clusters for $R < 3'$.  Finally, since the limiting magnitude $T_1 = 21.4$
will include $\simeq 78$\% of the total cluster population over all
luminosities for a GCLF dispersion of 1.3 mag, we derive
a final, {\sl very approximate} estimate $N_{GC} = 1550$ clusters.  This value
is unlikely to be accurate to better than $\pm 25$\%, but it suggests to us
that the estimate of $\simeq 980$ clusters over all radii 
by HHG04 was too conservative, and was probably 
the result of a slightly too-high background count level 
in the presence of the severe field contamination.
For a galaxy luminosity $M_V^T = -22.1$ (HHG04), the specific
frequency of the NGC 5128 cluster system is then $S_N \simeq 2.2 \pm 0.6$,
on the low end of the ``normal'' range for giant E galaxies.

Our new estimate of the cluster system size and profile obviously
needs much refinement.  Nevertheless, we believe that several hundred more
clusters in the halo of NGC 5128 remain to be found, by appropriately
designed search techniques.  Most of these will be in the
magnitude range $19.4 < T_1 < 21.4$ (within a magnitude of the
turnover point), corresponding roughly to $V \simeq 20 - 22$.
Furthermore, our surveys to this point have covered only the radial
region out to a $\sim 25'$ radius, which is less than 30 kpc from
the galaxy center.  It is already well known that the halo of NGC 5128
extends much further out than that \citep{peng04a,rej05}, indicating that significant
numbers of clusters at larger radii can probably be found as well.
Extrapolating the density curve in Fig.~\ref{fig7} outward to (say)
$R=50'$ (57 kpc) would predict anywhere from 100 to 400 more
clusters in the outer halo, depending on various plausible assumptions
about how steeply the GCS profile drops off at these larger radii
(see also HHG04).
Thus the global GC population in NGC 5128 over its entire halo 
might approach 2000.

The fact that such a rough, first-order argument
as we have just used can still yield a competitive estimate of the cluster
population is a forceful indicator of just how much we need a wide-field
imaging survey of the NGC 5128 halo under subarcsecond seeing conditions.

\section{Summary}

We have used deep HST/ACS images of 12 fields scattered around the bulge and
halo of NGC 5128 to find a total of 131 globular clusters or cluster candidates.
Of these, 62 are previously known GCs, leaving 69 newly discovered objects.
The objects range in magnitude from $T_1 = 16.5$ down to $T_1 = 22.7$.

The objects in our list are all highly probable globular clusters, with profiles
distinctly more extended than those of foreground stars.  Even with
first-order profile measurement (i.e.~rough FWHM values), they display a
wide range of magnitude and effective radius.  The luminosity distribution
(GCLF) for the clusters appears to be very close to normal for giant E galaxies,
once a small bright-end bias in the sample is corrected for.

Finally, a new estimate of the cluster population around NGC 5128 suggests that
the system contains $\sim 1500$ globular clusters over all magnitudes and
within a projected radius of $25'$.  If this is the case, then the specific
frequency of the system is $S_N \sim 2.2$.

\acknowledgements

WEH and GLHH thank the Natural Sciences and Engineering Research Council of Canada for
financial support.  DAF thanks the Australian Research Council for
financial support.  We are indebted to Mike Beasley for transmitting the velocity
data in the AAT cluster sample in advance of publication.

\clearpage

\begin{deluxetable}{lccrrrccrrc}
\rotate{}
\tablecaption{ACS/WFC Data for Previously Known NGC 5128 Clusters \label{clold}}
\tablewidth{0pt}
\tablehead{
\colhead{Cluster} & \colhead{$\alpha$} & \colhead{$\delta$} &
\colhead{$R'$} & \colhead{$\theta^o$} & \colhead{$T_1$} &
\colhead{$(M-T_1)$} & \colhead{$(C-T_1)$} & \colhead{$e$} & \colhead{FWHM} &
\colhead{Field} \\
}
\startdata
  PFF011         & 13 24 36.85 &-43 19 16.3 & 20.36 & 207.1 & 18.553 &  0.778 &  1.398 &  0.09 &   4.13 &C29      \\
  C29            & 13 24 40.37 &-43 18  5.3 & 19.02 & 207.0 & 17.533 &  0.838 &  1.924 &  0.08 &   3.92 &C29      \\
  PFF016         & 13 24 43.59 &-42 53  7.2 & 11.37 & 314.9 & 19.334 &  0.876 &  1.853 &  0.01 &   4.36 &C30      \\
  PFF021         & 13 24 54.16 &-42 54 50.4 &  8.78 & 315.9 & 18.814 &  0.703 &  1.318 &  0.03 &   3.14 &C3,C30   \\
  C30            & 13 24 54.35 &-42 53 24.7 &  9.84 & 321.8 & 16.681 &  0.813 &  1.789 &  0.05 &   5.71 &C30      \\
  PFF023         & 13 24 54.54 &-42 48 58.6 & 13.59 & 333.6 & 18.971 &  0.768 &  1.511 &  0.06 &   3.70 &C32      \\
  AAT111563      & 13 24 56.06 &-43 10 16.4 & 10.80 & 212.3 & 20.049 &  0.678 &  1.091 &  0.12 &   4.20 &C12      \\
  C3             & 13 24 58.21 &-42 56 10.0 &  7.33 & 312.8 & 17.081 &  0.801 &  1.940 &  0.06 &   3.94 &C3       \\
  PFF029         & 13 25  1.59 &-42 54 40.8 &  8.03 & 323.7 & 19.098 &  0.401 &  0.621 &  0.18 &   6.08 &C3       \\
  C4             & 13 25  1.81 &-43  9 25.5 &  9.53 & 209.7 & 17.498 &  0.738 &  1.451 &  0.15 &   6.10 &C12      \\
  PFF031         & 13 25  2.74 &-43 11 21.3 & 11.18 & 204.0 & 18.952 &  0.796 &  1.479 &  0.02 &   3.37 &C12      \\
  PFF034         & 13 25  3.35 &-43 11 39.7 & 11.41 & 202.9 & 19.384 &  0.808 &  1.502 &  0.07 &   4.31 &C12      \\
  C32            & 13 25  3.37 &-42 50 46.1 & 11.29 & 336.9 & 17.854 &  0.852 &  2.006 &  0.02 &   2.99 &C32      \\
  PFF035         & 13 25  4.45 &-43 10 48.5 & 10.55 & 203.7 & 19.132 &  0.936 &  2.000 &  0.04 &   4.70 &C12      \\
  C43            & 13 25  4.79 &-43  9 38.9 &  9.47 & 206.1 & 18.068 &  0.780 &  1.521 &  0.16 &   4.69 &C12      \\
  C12            & 13 25  5.69 &-43 10 30.8 & 10.19 & 203.2 & 17.358 &  0.869 &  1.984 &  0.17 &   4.15 &C12      \\
  WHH09          & 13 25  8.51 &-43  2 57.4 &  3.93 & 242.6 & 18.300 &  0.868 &  1.979 &  0.02 &   4.63 &C6       \\
  f2.GC70        & 13 25  8.93 &-43  8 53.7 &  8.47 & 203.8 & 20.039 &  0.663 &  1.253 &  0.19 &   7.21 &C12      \\
  f2.GC69        & 13 25  9.06 &-43 10  2.1 &  9.51 & 200.9 & 19.301 &  0.917 &  1.813 &  0.03 &   2.87 &C12      \\
  AAT113992      & 13 25 10.49 &-43  3 24.0 &  3.86 & 234.3 & 19.818 &  0.966 &  1.955 &  0.03 &   4.85 &C6       \\
  C14            & 13 25 10.49 &-42 44 52.8 & 16.57 & 349.1 & 17.407 &  0.752 &  1.655 &  0.07 &   4.38 &C14      \\
  PFF041         & 13 25 11.14 &-43  3  9.6 &  3.62 & 236.2 & 19.007 &  0.761 &  1.471 &  0.02 &   3.40 &C6       \\
  AAT115339      & 13 25 18.42 &-43  4  9.8 &  3.45 & 209.1 & 19.561 &  0.815 &  1.546 &  0.04 &   3.67 &C6       \\
  C6             & 13 25 22.17 &-43  2 45.9 &  1.90 & 211.6 & 16.510 &  0.835 &  1.858 &  0.17 &   5.61 &C6       \\
  PFF052         & 13 25 25.73 &-43  5 16.6 &  4.14 & 184.8 & 19.416 &  0.755 &  1.442 &  0.05 &   3.59 &C18      \\
  WHH16/K102     & 13 25 27.97 &-43  4  2.2 &  2.89 & 178.7 & 18.599 &  0.925 &  2.036 &  0.04 &   3.60 &C18      \\
  f2.GC31        & 13 25 29.45 &-43  7 41.7 &  6.56 & 177.1 & 20.154 &  0.772 &  1.395 &  0.10 &   4.02 &C19      \\
  f2.GC28        & 13 25 30.17 &-43  6 54.6 &  5.78 & 175.4 & 21.156 &  0.663 &  1.701 &  0.05 &   7.40 &C19      \\
  AAT117287      & 13 25 31.06 &-43  4 16.9 &  3.20 & 168.6 & 20.450 &  0.107 &  1.374 &  0.02 &   6.87 &C18      \\
  f2.GC23        & 13 25 32.78 &-43  7  2.3 &  5.97 & 170.9 & 18.186 &  0.789 &  1.708 &  0.02 &   3.22 &C19      \\
  f2.GC20        & 13 25 33.14 &-43  7  1.3 &  5.96 & 170.2 & 21.013 &  0.766 &  1.870 &  0.14 &   5.46 &C19      \\
  f2.GC18        & 13 25 33.65 &-43  7 19.2 &  6.27 & 169.9 & 21.147 &  0.781 &  \ldots &  0.02 &   3.53 &C19      \\
  K131           & 13 25 32.86 &-43  4 29.1 &  3.47 & 164.0 & 18.741 &  0.906 &  2.104 &  0.02 &   2.76 &C18      \\
  WHH22          & 13 25 35.31 &-43  5 29.0 &  4.56 & 162.0 & 18.032 &  0.766 &  1.631 &  0.03 &   3.50 &C18,C19  \\
  f2.GC14        & 13 25 35.89 &-43  7 16.1 &  6.31 & 166.1 & 20.824 &  0.439 &  1.678 &  0.09 &   4.06 &C19      \\
  AAT118198      & 13 25 37.45 &-43  5 45.0 &  4.94 & 158.7 & 19.031 &  0.909 &  2.110 &  0.06 &   2.79 &C18,C19  \\
  PFF059         & 13 25 39.63 &-43  4  1.4 &  3.62 & 142.6 & 19.479 &  0.855 &  1.943 &  0.04 &   4.36 &C18      \\
  C18/K163       & 13 25 39.86 &-43  5  1.8 &  4.48 & 150.0 & 16.891 &  0.618 &  1.603 &  0.17 &   4.79 &C18,C19  \\
  f2.GC03        & 13 25 40.89 &-43  8 16.1 &  7.52 & 161.2 & 19.384 &  0.700 &  1.240 &  0.04 &   3.33 &C19      \\
  AAT119508      & 13 25 42.51 &-43  3 41.5 &  3.73 & 133.1 & 19.857 &  0.883 &  1.832 &  0.12 &   6.46 &C18      \\
  C19            & 13 25 43.38 &-43  7 22.9 &  6.87 & 155.2 & 17.554 &  0.780 &  1.661 &  0.03 &   4.64 &C19      \\
  PFF063         & 13 25 43.79 &-43  7 55.0 &  7.39 & 156.4 & 19.452 &  0.674 &  1.176 &  0.08 &   3.15 &C19      \\
  G284           & 13 25 46.58 &-42 57  3.0 &  5.37 &  40.2 & 19.412 &  0.789 &  1.853 &  0.04 &   2.85 &C25      \\
  PFF066         & 13 25 47.14 &-43  6  8.9 &  6.14 & 144.5 & 19.436 &  0.774 &  1.629 &  0.05 &   3.89 &C19      \\
  AAT120336      & 13 25 50.21 &-43  6  8.6 &  6.48 & 140.4 & 19.668 &  0.835 &  1.809 &  0.07 &   4.74 &C19      \\
  AAT120976      & 13 25 54.28 &-42 56 20.5 &  6.84 &  45.4 & 19.991 &  0.740 &  1.511 &  0.03 &   4.19 &C7,C25   \\
  PFF079         & 13 25 58.90 &-42 53 18.8 &  9.70 &  36.1 & 19.140 &  0.718 &  1.484 &  0.04 &   3.05 &C37      \\
  C104/AAT121826 & 13 25 59.47 &-42 55 30.7 &  8.10 &  45.9 & 19.403 &  0.743 &  1.405 &  0.08 &   3.58 &C7,C25   \\
  G221           & 13 26  1.10 &-42 55 13.4 &  8.52 &  45.9 & 18.828 &  0.775 &  1.715 &  0.06 &   3.75 &C7,C25   \\
  PFF083         & 13 26  1.81 &-42 58 15.0 &  6.89 &  65.1 & 19.436 &  0.774 &  1.683 &  0.04 &   3.17 &C25      \\
  C25            & 13 26  2.84 &-42 56 57.0 &  7.68 &  56.9 & 17.965 &  0.863 &  1.952 &  0.13 &   3.44 &C7,C25   \\
  G293           & 13 26  4.19 &-42 55 44.7 &  8.59 &  51.1 & 18.693 &  0.664 &  1.349 &  0.07 &   3.28 &C7,C25   \\
  C105           & 13 26  5.04 &-42 55 36.3 &  8.80 &  51.0 & 21.758 &  0.821 &  1.922 &  0.32 &  10.38 &C25      \\
  f1.GC20        & 13 26  5.38 &-42 55 22.4 &  9.00 &  50.1 & 21.216 &  0.788 &  1.839 &  0.26 &  10.39 &C7,C25   \\
  C7             & 13 26  5.40 &-42 56 32.4 &  8.30 &  56.3 & 16.644 &  0.682 &  1.534 &  0.08 &   5.83 &C7,C25   \\
  G170           & 13 26  6.92 &-42 57 35.0 &  8.02 &  63.6 & 18.727 &  0.835 &  1.844 &  0.03 &   3.72 &C7       \\
  C36            & 13 26  7.72 &-42 52  0.2 & 11.72 &  38.7 & 17.944 &  0.667 &  1.378 &  0.14 &   3.57 &C37      \\
  f1.GC15        & 13 26  8.87 &-42 53 42.5 & 10.59 &  45.4 & 19.531 &  0.915 &  2.224 &  0.07 &   3.36 &C37      \\
  f1.GC14        & 13 26  9.37 &-42 53 17.5 & 10.95 &  44.2 & 19.671 &  0.712 &  1.421 &  0.06 &   6.95 &C37      \\
  f1.GC34        & 13 26  9.67 &-42 53 16.9 & 11.00 &  44.3 & 20.961 &  0.855 &  1.896 &  0.03 &   3.15 &C37      \\
  f1.GC21        & 13 26 10.51 &-42 55  0.2 &  9.96 &  51.9 & 21.358 &  0.956 &  2.071 &  0.02 &   6.83 &C37      \\
  C37            & 13 26 10.57 &-42 53 42.6 & 10.81 &  46.6 & 17.962 &  0.758 &  1.691 &  0.03 &   3.15 &C37      \\
\enddata

\end{deluxetable}

\begin{deluxetable}{lccrrrccrrcl}
\rotate{}
\tablecaption{ACS/WFC Data for New NGC 5128 Clusters \label{clnew}}
\tablewidth{0pt}
\tablehead{
\colhead{Cluster} & \colhead{$\alpha$} & \colhead{$\delta$} &
\colhead{$R'$} & \colhead{$\theta^o$} & \colhead{$T_1$} &
\colhead{$(M-T_1)$} & \colhead{$(C-T_1)$} & \colhead{$e$} & \colhead{FWHM} &
\colhead{Field} & \colhead{Notes} \\
}
\startdata
  C111 & 13 24 26.97 &-43 17 20.0 & 19.62 & 214.4 & 21.363 &  0.885 &  1.172 &  0.03 &   4.23 &C29      \\
  C112 & 13 24 32.66 &-43 18 48.8 & 20.32 & 209.6 & 21.310 &  0.913 &  1.668 &  0.21 &   9.79 &C29      \\
  C113 & 13 24 37.75 &-43 16 26.5 & 17.81 & 210.8 & 19.120 &  0.748 &  1.378 &  0.13 &   3.30 &C29      \\
  C114 & 13 24 40.48 &-42 53 35.3 & 11.46 & 311.3 & 21.423 &  0.971 &  2.037 &  0.12 &  12.96 &C30      \\
  C115 & 13 24 48.71 &-42 52 35.5 & 11.12 & 320.3 & 19.507 &  0.726 &  1.372 &  0.05 &   5.62 &C30      \\
  C116 & 13 24 55.46 &-43  9 58.5 & 10.61 & 213.7 & 21.837 &  0.901 &  1.967 &  0.12 &   5.34 &C12      \\
  C117 & 13 24 56.06 &-42 54 29.6 &  8.81 & 319.1 & 19.262 &  0.920 &  2.011 &  0.07 &   3.75 &C3       \\
  C118 & 13 24 57.17 &-43  8 42.6 &  9.39 & 216.3 & 20.672 &  1.085 &  1.924 &  0.30 &   3.98 &C12      \\
  C119 & 13 24 57.69 &-42 55 48.4 &  7.64 & 314.3 & 20.669 &  0.349 &  0.487 &  0.11 &   7.09 &C3       \\
  C120 & 13 24 57.95 &-42 52  4.9 & 10.56 & 329.1 & 21.433 &  0.904 &  1.828 &  0.08 &  15.64 &C30      \\
  C121 & 13 24 58.42 &-43  8 21.2 &  8.97 & 216.5 & 22.446 &  0.565 &  1.080 &  0.22 &   5.79 &C12      \\
  C122 & 13 24 59.01 &-43  8 21.4 &  8.91 & 215.9 & 22.330 &  \ldots & \ldots  &  0.10 &   5.30 &C12 & (1)\\
  C123 & 13 24 59.92 &-43  9  8.6 &  9.46 & 212.3 & 20.244 &  0.859 &  1.637 &  0.10 &   4.05 &C12      \\
  C124 & 13 25  0.37 &-43 10 46.9 & 10.85 & 207.3 & 21.574 &  0.784 &  1.498 &  0.05 &   6.61 &C12      \\
  C125 & 13 25  0.83 &-43 11 10.6 & 11.16 & 206.0 & 20.664 &  0.887 &  1.671 &  0.08 &   6.07 &C12      \\
  C126 & 13 25  0.91 &-43  9 14.5 &  9.45 & 211.1 & 22.663 &  0.875 &  1.354 &  0.19 &   4.08 &C12      \\
  C127 & 13 25  1.32 &-43  8 43.4 &  8.97 & 212.4 & 21.620 &  0.801 &  1.593 &  0.31 &   6.13 &C12      \\
  C128 & 13 25  1.46 &-43  8 33.0 &  8.81 & 212.9 & 20.947 &  1.089 &  2.131 &  0.07 &  10.23 &C12      \\
  C129 & 13 25  1.63 &-42 50 51.3 & 11.34 & 335.2 & 20.831 &  0.876 &  1.757 &  0.03 &   4.45 &C32      \\
  C130 & 13 25  1.86 &-42 52 27.8 &  9.88 & 331.5 & 19.851 &  0.730 &  1.368 &  0.09 &   3.69 &C30      \\
  C131 & 13 25  3.67 &-42 51 21.7 & 10.72 & 335.9 & 20.202 &  0.778 &  1.363 &  0.08 &   5.87 &C32      \\
  C132 & 13 25  8.79 &-43  9  9.6 &  8.72 & 203.2 & 18.896 &  0.794 &  1.480 &  0.06 &   3.60 &C12      \\
  C133 & 13 25 11.05 &-43  1 32.3 &  3.05 & 262.6 & 19.300 &  \ldots &  \ldots &  0.12 &   3.80 &C6  & (3)\\
  C134 & 13 25 13.20 &-43  2 31.3 &  2.97 & 242.4 & 20.711 &  0.931 &  1.735 &  0.34 &   6.11 &C6       \\
  C135 & 13 25 14.07 &-43  0 51.8 &  2.49 & 276.5 & 18.900 &  \ldots &  \ldots &  0.11 &   3.56 &C6  & (3)\\
  C136 & 13 25 14.07 &-43  3 35.0 &  3.47 & 225.4 & 21.190 &  0.305 &  1.295 &  0.11 &   4.90 &C6       \\
  C137 & 13 25 16.06 &-43  2 19.3 &  2.42 & 240.9 & 19.040 &  \ldots &  \ldots &  0.03 &   3.69 &C6  & (3)\\
  C138 & 13 25 16.91 &-43  3  8.0 &  2.79 & 224.6 & 19.974 &  0.815 &  1.555 &  0.03 &   3.80 &C6       \\
  C139 & 13 25 17.06 &-43  2 44.6 &  2.50 & 230.4 & 18.863 &  0.760 &  1.866 &  0.03 &   3.26 &C6       \\
  C140 & 13 25 17.42 &-43  3 25.2 &  2.94 & 219.3 & 19.830 &  0.818 &  1.570 &  0.06 &   5.01 &C6       \\
  C141 & 13 25 18.14 &-43  2 50.9 &  2.43 & 225.5 & 20.945 &  0.604 &  1.288 &  0.60 &  12.11 &C6       \\
  C142 & 13 25 18.50 &-43  1 16.4 &  1.67 & 265.7 & 17.640 &  \ldots &  \ldots &  0.09 &   4.41 &C6       \\
  C143 & 13 25 23.20 &-43  3 12.9 &  2.22 & 201.3 & 20.525 &  0.823 &  1.345 &  0.22 &   5.83 &C6       \\
  C144 & 13 25 26.28 &-43  4 38.5 &  3.50 & 184.0 & 21.959 &  0.405 &  1.257 &  0.68 &   8.76 &C18      \\
  C145 & 13 25 28.81 &-43  4 21.6 &  3.22 & 176.1 & 17.814 &  0.726 &  1.510 &  0.06 &   3.99 &C18      \\
  C146 & 13 25 29.87 &-43  5  9.2 &  4.03 & 174.1 & 19.940 &  0.843 &  1.773 &  0.04 &   4.08 &C18      \\
  C147 & 13 25 30.65 &-43  3 47.1 &  2.70 & 168.1 & 20.055 &  0.829 &  1.523 &  0.12 &   5.77 &C18      \\
  C148 & 13 25 31.75 &-43  5 46.0 &  4.68 & 170.7 & 20.207 &  0.585 &  1.042 &  0.16 &  10.55 &C18      \\
  C149 & 13 25 32.32 &-43  7 17.1 &  6.20 & 172.0 & 19.680 &  0.743 &  1.343 &  0.03 &   3.67 &C19      \\
  C150 & 13 25 33.82 &-43  2 49.6 &  2.03 & 146.0 & 19.780 &  \ldots &  \ldots &  0.11 &   3.75 &C18 & (3)\\
  C151 & 13 25 33.93 &-43  3 51.4 &  2.94 & 156.9 & 19.952 &  0.948 &  2.317 &  0.02 &   3.42 &C18      \\
  C152 & 13 25 34.64 &-43  3 16.4 &  2.48 & 148.9 & 17.820 &  \ldots &  \ldots &  0.10 &   4.80 &C18 & (3)\\
  C153 & 13 25 34.64 &-43  3 27.8 &  2.65 & 151.0 & 18.230 &  \ldots &  \ldots &  0.03 &   5.10 &C18 & (3)\\
  C154 & 13 25 34.71 &-43  3 30.2 &  2.69 & 151.2 & 19.580 &  \ldots &  \ldots &  0.08 &   4.48 &C18 & (3)\\
  C155 & 13 25 36.47 &-43  8  3.5 &  7.10 & 166.8 & 21.358 &  0.855 &  1.627 &  0.28 &   9.20 &C19      \\
  C156 & 13 25 38.43 &-43  5  2.6 &  4.37 & 153.1 & 17.651 &  0.899 &  2.106 &  0.10 &   3.66 &C18,C19  \\
  C157 & 13 25 38.45 &-43  3 28.9 &  3.06 & 139.7 & 19.210 &  \ldots & \ldots &  0.03 &   3.77 &C18 & (2)  \\
  C158 & 13 25 38.76 &-43  5 34.5 &  4.88 & 155.3 & 20.061 &  0.801 &  1.606 &  0.04 &   7.41 &C18,C19  \\
  C159 & 13 25 39.17 &-43  4 33.8 &  4.02 & 148.3 & 19.929 &  0.896 &  1.986 &  0.03 &   3.30 &C18      \\
  C160 & 13 25 40.09 &-43  3  7.1 &  3.01 & 130.8 & 19.990 &  \ldots &  \ldots &  0.18 &   4.20 &C18 & (2)\\
  C161 & 13 25 40.52 &-43  7 17.9 &  6.59 & 159.0 & 19.261 &  0.892 &  1.953 &  0.00 &   3.28 &C19      \\
  C162 & 13 25 40.87 &-43  5  0.4 &  4.56 & 147.9 & 20.909 &  0.505 &  1.046 &  0.19 &   9.15 &C18      \\
  C163 & 13 25 41.63 &-43  3 45.8 &  3.66 & 135.6 & 20.431 &  0.815 &  1.547 &  0.13 &   8.06 &C18      \\
  C164 & 13 25 42.09 &-43  3 19.5 &  3.43 & 129.5 & 19.600 &  \ldots &  \ldots &  0.08 &   4.22 &C18 & (4)\\
  C165 & 13 25 43.43 &-43  4 56.5 &  4.77 & 142.7 & 18.173 &  0.812 &  1.931 &  0.00 &   3.71 &C18      \\
  C166 & 13 25 44.90 &-43  4 21.1 &  4.50 & 135.4 & 20.720 &  \ldots &  \ldots &  0.29 &  12.46 &C18 & (1)\\
  C167 & 13 25 45.97 &-43  6 45.4 &  6.54 & 149.1 & 20.410 &  \ldots &  \ldots &  0.05 &   3.59 &C19 & (3)\\
  C168 & 13 25 48.46 &-43  7 12.5 &  7.16 & 147.9 & 19.710 &  \ldots &  \ldots &  0.04 &   4.30 &C19 & (5)\\
  C169 & 13 25 51.01 &-42 55 36.3 &  7.00 &  37.7 & 20.389 &  0.616 &  1.212 &  0.13 &   6.27 &C25      \\
  C170 & 13 25 56.11 &-42 56 12.9 &  7.17 &  46.6 & 22.161 &  0.670 &  1.148 &  0.19 &   4.36 &C25      \\
  C171 & 13 25 57.78 &-42 55 36.1 &  7.82 &  44.8 & 20.670 &  0.784 &  1.577 &  0.08 &  11.31 &C7,C25   \\
  C172 & 13 25 57.95 &-42 53  4.3 &  9.80 &  34.5 & 20.907 &  0.546 &  1.242 &  0.04 &   5.76 &C37      \\
  C173 & 13 25 59.57 &-42 55  1.5 &  8.46 &  43.7 & 21.057 &  0.945 &  2.092 &  0.04 &   9.40 &C7,C25   \\
  C174 & 13 25 59.63 &-42 55 15.7 &  8.30 &  44.8 & 21.424 &  0.992 &  1.812 &  0.04 &   4.66 &C25      \\
  C175 & 13 26  0.93 &-42 58 28.9 &  6.65 &  66.4 & 21.838 &  0.694 &  1.668 &  0.00 &   5.81 &C25      \\
  C176 & 13 26  2.79 &-42 57  5.0 &  7.61 &  57.7 & 21.297 &  0.621 &  1.053 &  0.13 &   5.32 &C7,C25   \\
  C177 & 13 26  3.20 &-42 54 30.1 &  9.30 &  44.4 & 21.087 &  1.024 &  1.867 &  0.24 &   7.01 &C7       \\
  C178 & 13 26  3.85 &-42 56 45.3 &  7.95 &  56.5 & 21.534 &  0.791 &  1.481 &  0.13 &   7.71 &C25      \\
  C179 & 13 26  9.87 &-42 56 36.0 &  8.96 &  59.5 & 21.043 &  0.672 &  1.088 &  0.09 &   8.24 &C7       \\
\enddata

Notes to Table 1:
(1) Faint globular?
(2) Very crowded, in inner bulge region
(3) Very compact globular or possible star
(4) Possible globular, slightly elliptical
(5) Globular next to very bright star

\end{deluxetable}

\begin{deluxetable}{lccrcc}
\tablecaption{Revised Coordinates for Clusters from Paper I \label{clstis}}
\tablewidth{0pt}
\tablehead{
\colhead{Cluster} & \colhead{$\alpha$} & \colhead{$\delta$} &
\colhead{$T_1$} & \colhead{$(M-T_1)$} & \colhead{$(C-T_1)$} \\
}
\startdata
C40      & 13 23 42.33 & -43 09 37.8 &  18.490 &   0.791 &   1.630 \\
C41      & 13 24 38.97 & -43 20 06.5 &  17.969 &   0.951 &   1.980 \\
C29      & 13 24 40.37 & -43 18 05.3 &  17.533 &   0.838 &   1.924 \\
G19      & 13 24 46.44 & -43 04 11.6 &  18.636 &   0.748 &   1.422 \\
G277     & 13 24 47.36 & -42 58 29.9 &  18.607 &   0.764 &   1.530 \\
C2       & 13 24 51.47 & -43 12 11.2 &  17.937 &   0.778 &   1.546 \\
C100     & 13 24 52.06 & -43 04 32.7 &  19.030 &   0.694 &   1.409 \\
G302     & 13 24 53.27 & -43 04 34.8 &  18.728 &   0.775 &   1.450 \\
C11      & 13 24 54.70 & -43 01 21.6 &  17.197 &   0.786 &   2.011 \\
C31      & 13 24 57.42 & -43 01 08.1 &  17.710 &   0.881 &   2.023 \\
C32      & 13 25 03.37 & -42 50 46.1 &  17.854 &   0.852 &   2.006 \\
C44      & 13 25 31.73 & -43 19 22.8 &  18.148 &   0.701 &   1.441 \\
C17      & 13 25 39.72 & -42 55 59.1 &  17.186 &   0.533 &   1.422 \\
C101     & 13 25 40.56 & -42 56 01.0 &  20.077 &   0.820 &   1.762 \\
C102     & 13 25 52.09 & -42 59 12.8 &  21.038 &   0.829 &   1.673 \\
C21      & 13 25 52.73 & -43 05 46.5 &  17.397 &   0.773 &   1.576 \\
C22      & 13 25 53.55 & -42 59 07.5 &  17.696 &   0.643 &   1.516 \\
C23      & 13 25 54.57 & -42 59 25.3 &  16.686 &   0.749 &   1.904 \\
C103     & 13 25 55.01 & -42 59 13.8 &  18.439 &   1.997 &   1.993 \\
C104     & 13 25 59.48 & -42 55 30.7 &  19.403 &   0.742 &   1.404 \\
G221     & 13 26 01.10 & -42 55 13.4 &  18.828 &   0.775 &   1.715 \\
C25      & 13 26 02.83 & -42 56 57.0 &  17.965 &   0.862 &   1.952 \\
G293     & 13 26 04.19 & -42 55 44.7 &  18.693 &   0.665 &   1.349 \\
C105     & 13 26 05.04 & -42 55 36.3 &  21.758 &   0.821 &   1.922 \\
C7       & 13 26 05.40 & -42 56 32.4 &  16.644 &   0.681 &   1.533 \\
C106     & 13 26 06.19 & -42 56 44.1 &  20.657 &   0.799 &   1.879 \\
C37      & 13 26 10.57 & -42 53 42.6 &  17.962 &   0.758 &   1.691 \\
\enddata

\end{deluxetable}
%
%

\begin{figure} 
\figurenum{2a} 
\plotone{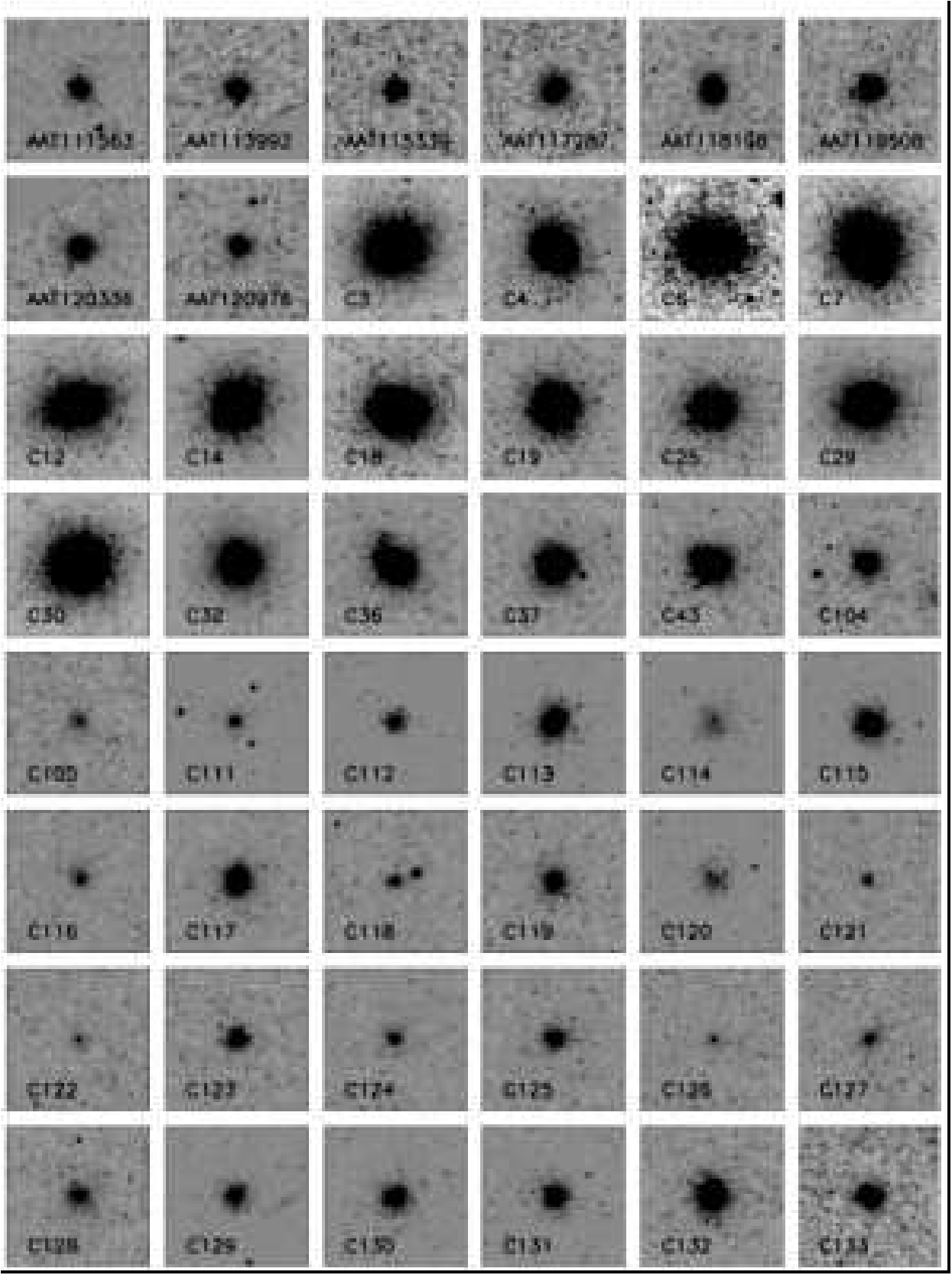} 
\caption{Images of individual globular clusters in NGC 5128.  
The field size of each image is $5''$ across,
equivalent to a linear distance of $\simeq 100$ pc at the distance
of NGC 5128.  North is upward and East to the left in all cases.
Identification labels for each cluster are from 
Tables 1 and 2.}
\label{fig2a} 
\end{figure}

\begin{figure} 
\figurenum{2b} 
\plotone{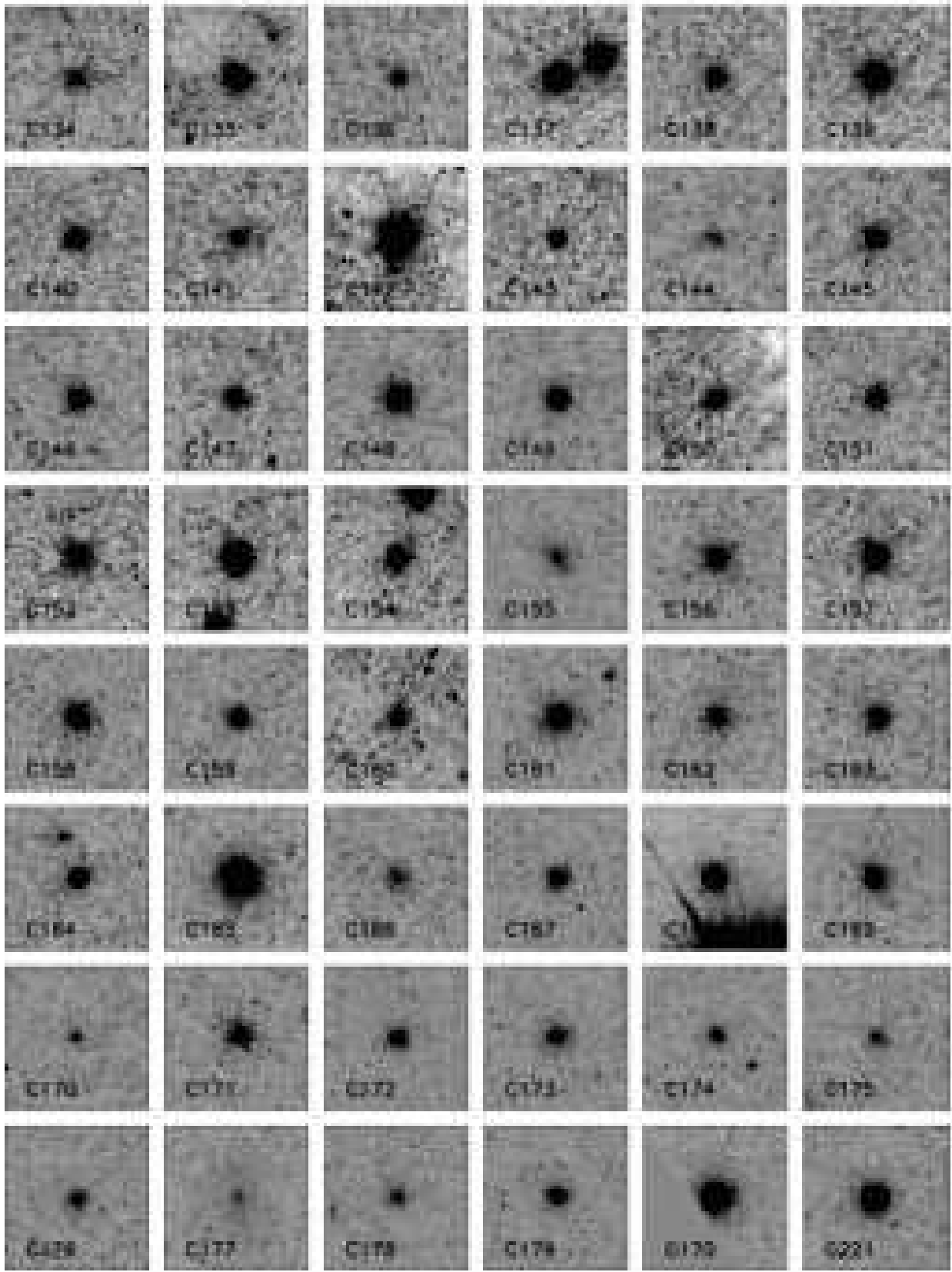} 
\caption{Images of a second group of individual globular clusters in NGC 5128.}
\label{fig2b} 
\end{figure}

\begin{figure} 
\figurenum{2c} 
\plotone{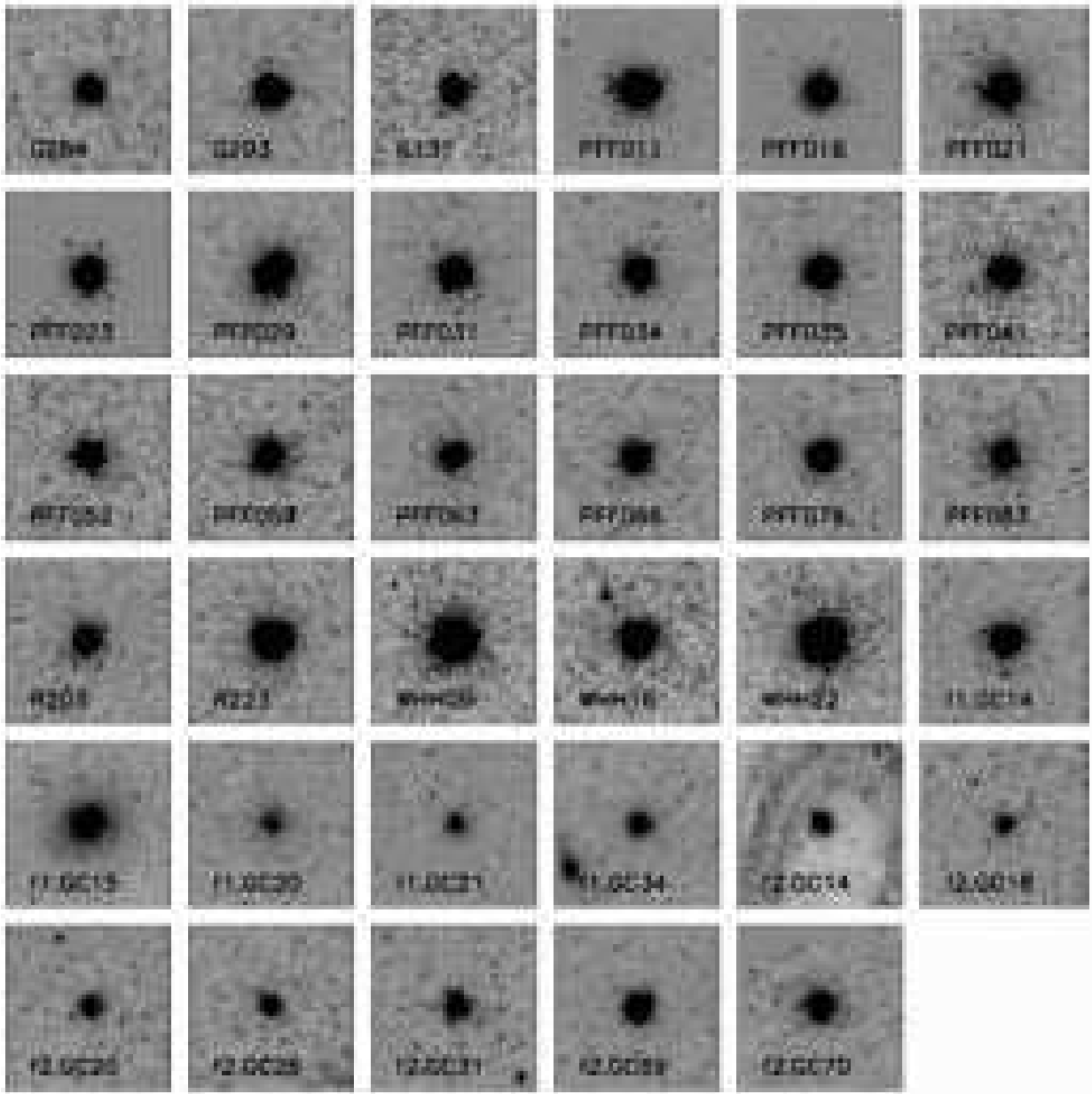} 
\caption{Images of the final group of individual globular clusters in NGC 5128.}
\label{fig2c} 
\end{figure}

\begin{figure} 
\figurenum{3} 
\plotone{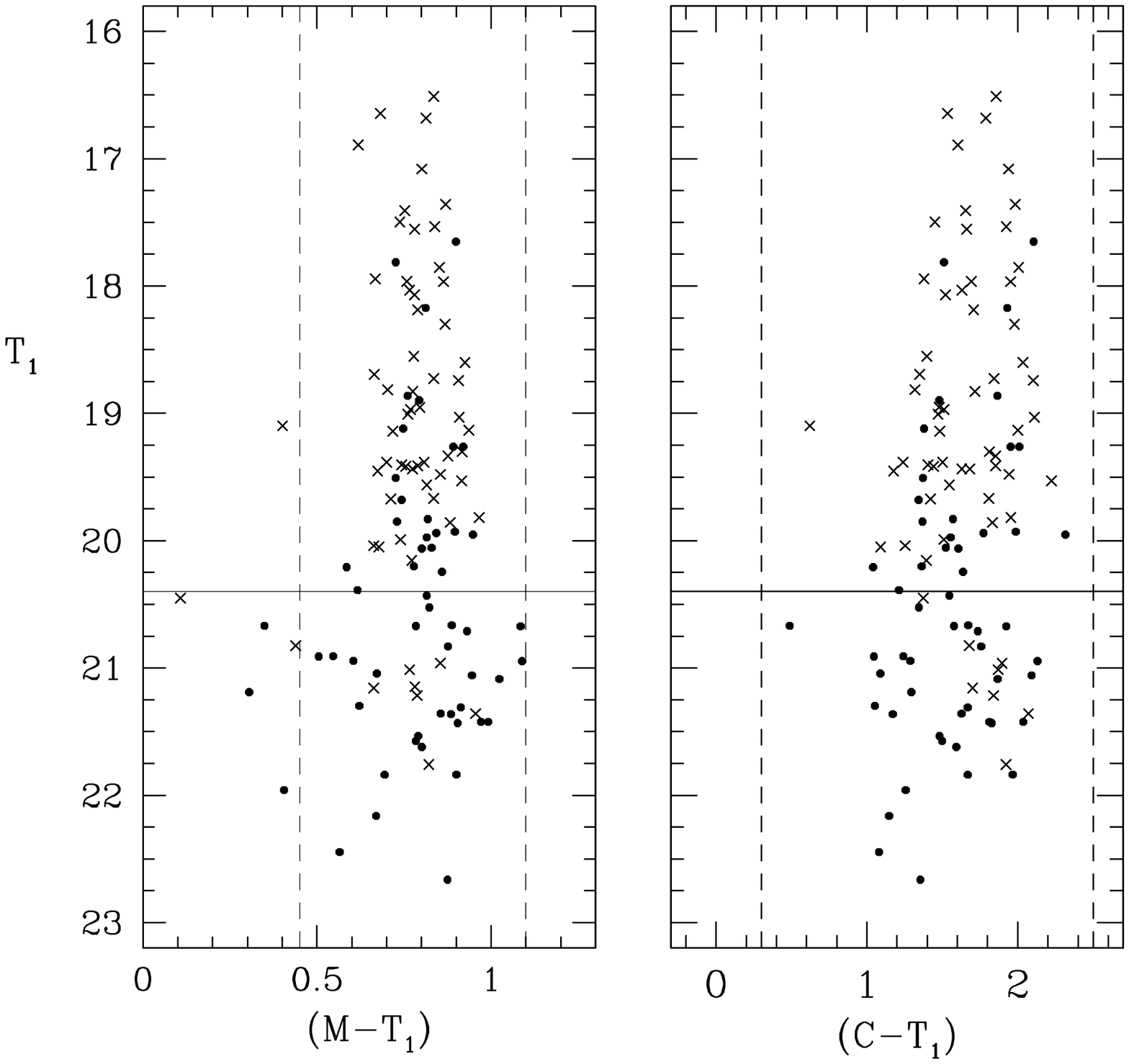} 
\caption{Color-magnitude data for the objects in Tables 1 -- 3.
The Washington color indices $(M-T_1)$ and $(C-T_1)$ are plotted
against magnitude $T_1 \simeq R$.  In each panel, the solid dots
are the newly identified cluster candidates from Table 2, while
the crosses are the previously known clusters (or cluster candidates)
from Tables 1 and 3.  The vertical dashed lines in each panel represent
generous blue and red color boundaries enclosing normal, old globular
clusters (see Harris et al. 2004), while the horizontal line
shows the level of the expected ``turnover point'' (maximum frequency) of the globular
cluster luminosity function.}
\label{fig2}
\end{figure}

\begin{figure} 
\figurenum{4} 
\plotone{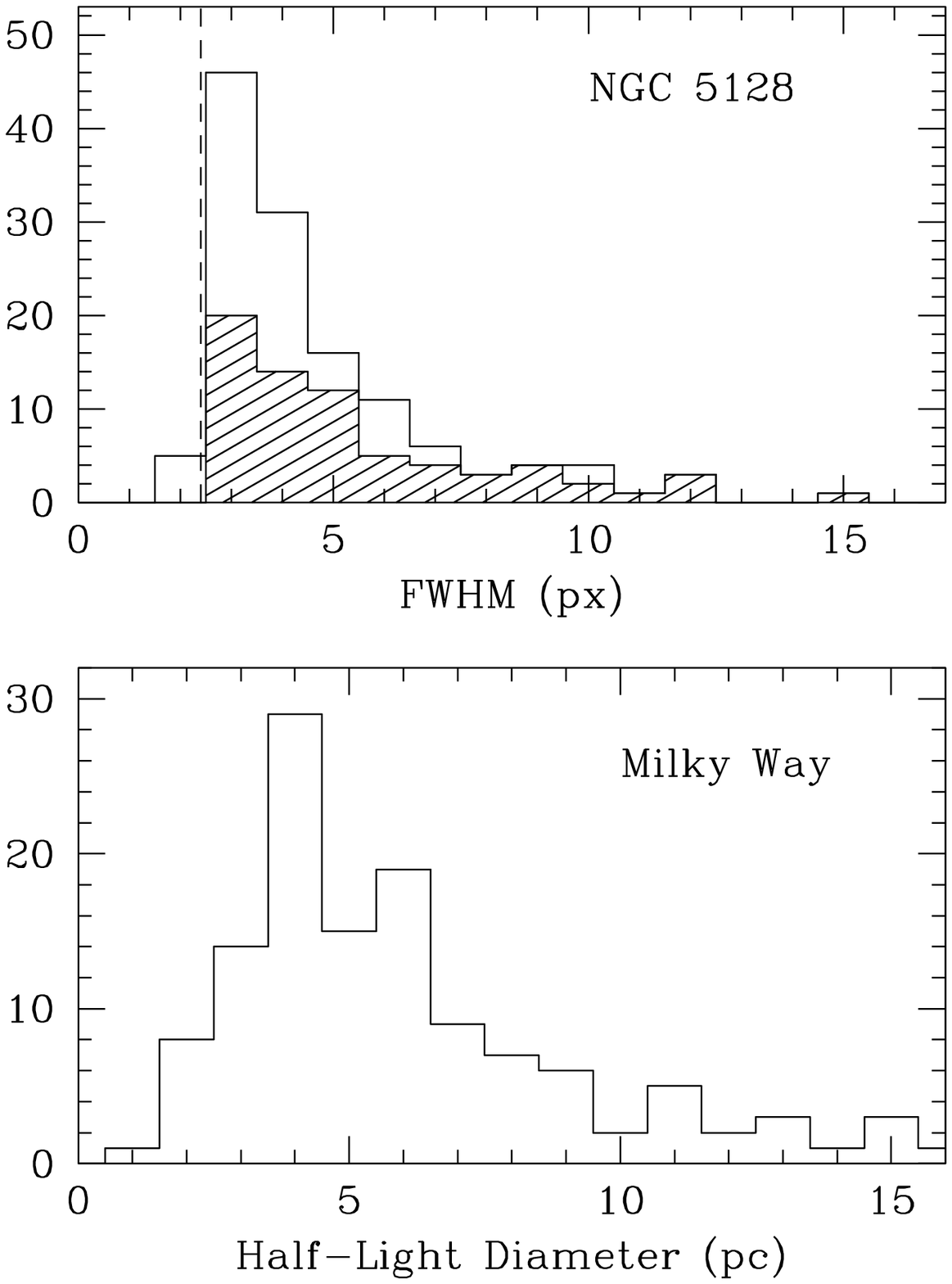} 
\caption{{\sl Upper panel:} Distribution of FWHM values for globular
clusters in NGC 5128.
The shaded region represents the new cluster candidates from Table 2,
while the unshaded region represents the previously known clusters from
Table 1.  The scale is approximately 1 pixel $\simeq 1$ parsec.  
The typical FWHM for the point spread function on our ACS/WFC images is
2.4 pixels, shown by the vertical dashed line.
{\sl Lower panel:}  Distribution of half-light diameters ($= 2 r_h$)
for globular clusters in the Milky Way, from Harris (1996).}
\label{fig3}
\end{figure}

\begin{figure} 
\figurenum{5} 
\plotone{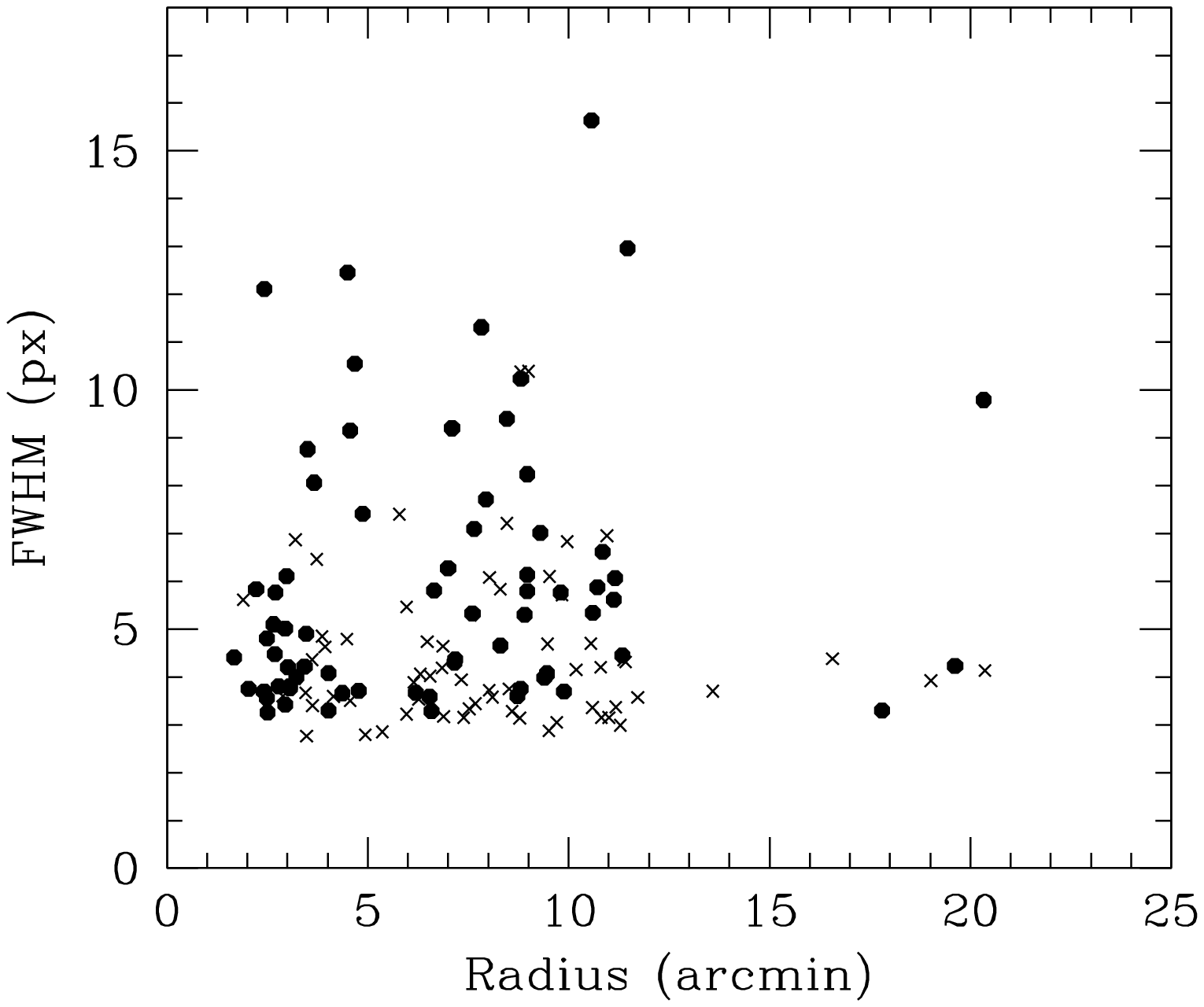} 
\caption{FWHM values for globular clusters on our ACS images, plotted
against projected radius from the center of NGC 5128.  Solid symbols are the newly
discovered clusters from Table 2, while crosses are the previously known
ones from Table 1.}
\label{fig4}
\end{figure}

\begin{figure} 
\figurenum{6} 
\plotone{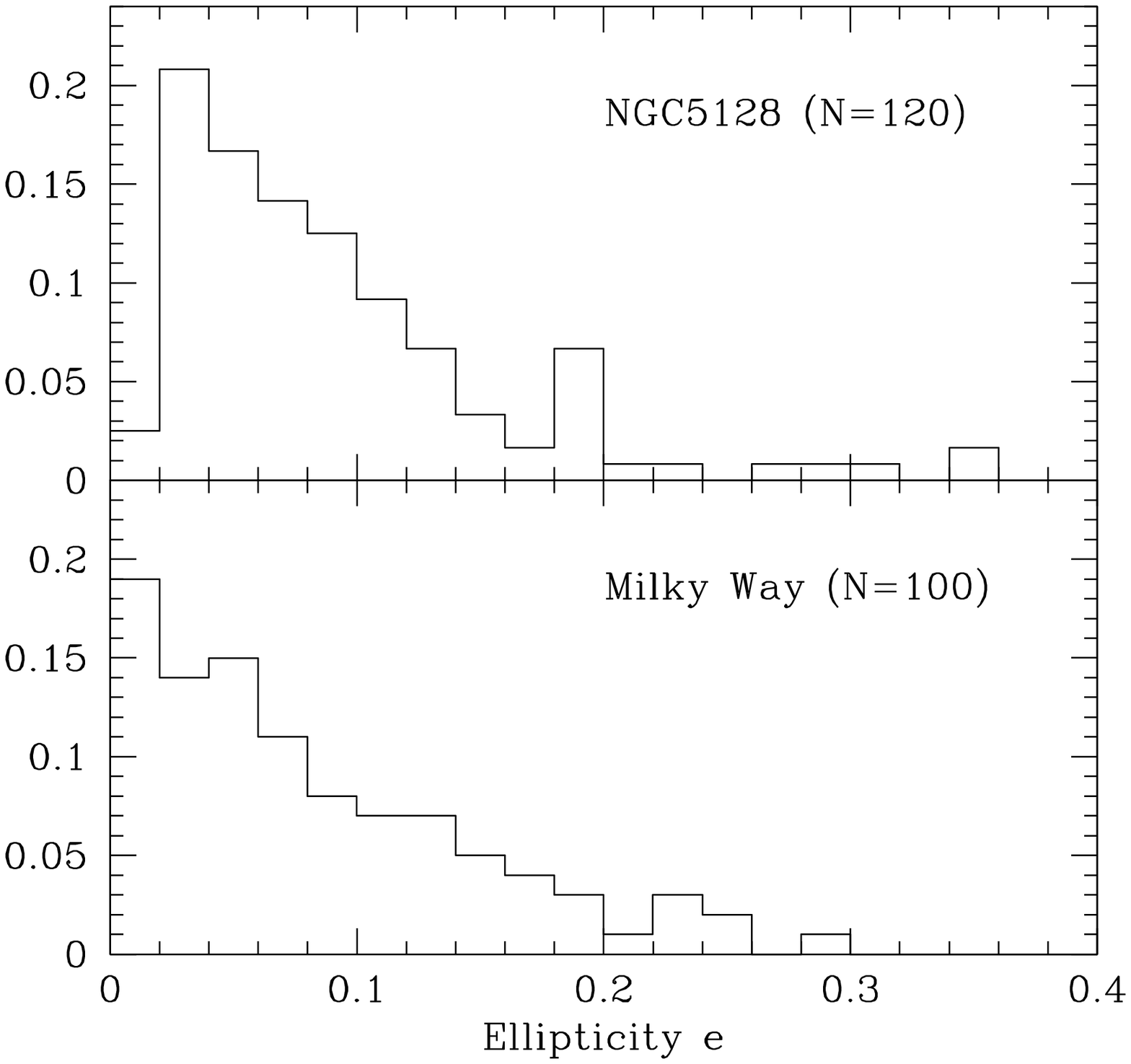} 
\caption{Histogram of ellipticities for 120 globular clusters in
NGC 5128 and 100 globular clusters in the Milky Way.}
\label{fig5}
\end{figure}

\begin{figure} 
\figurenum{7} 
\plotone{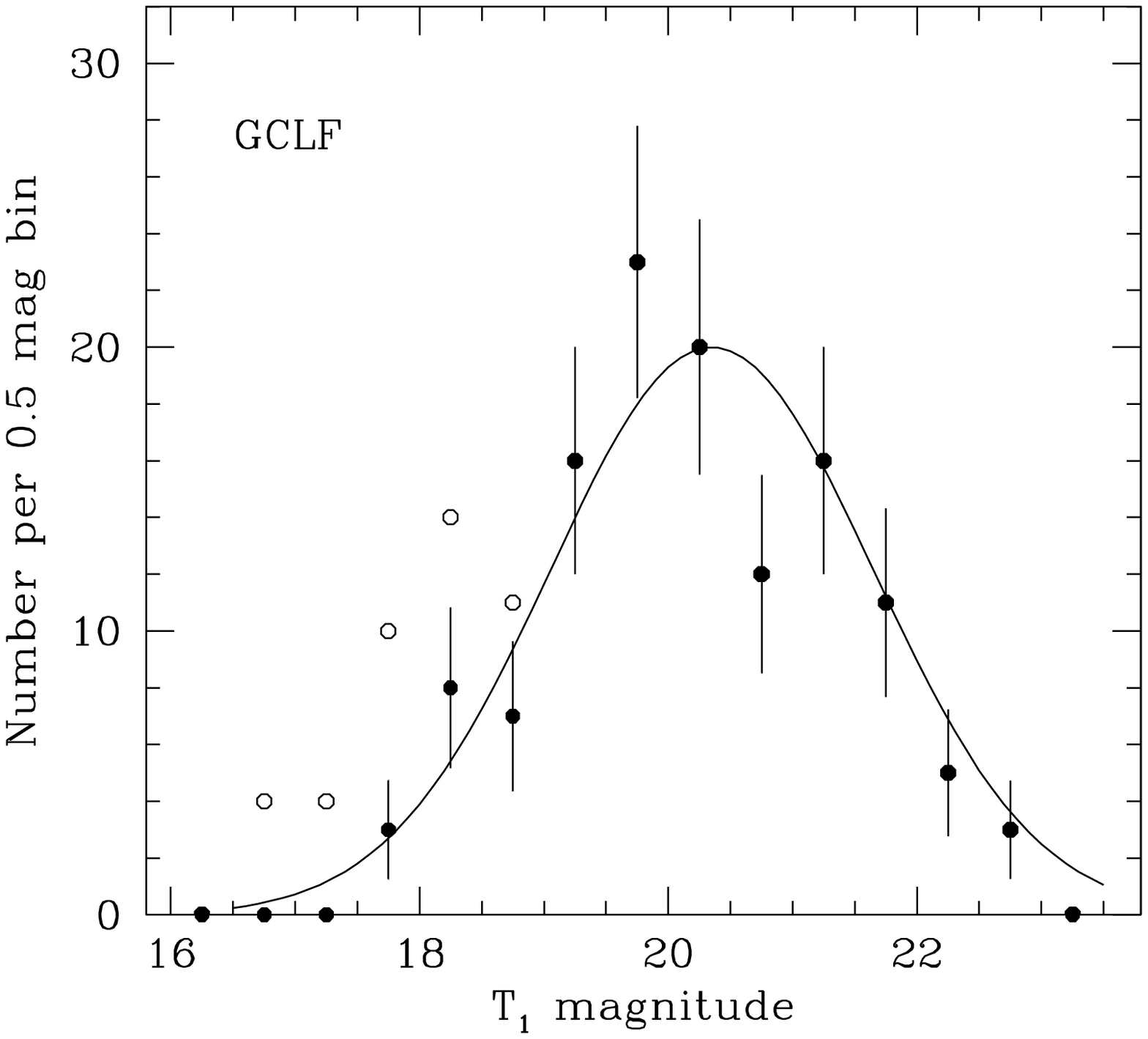} 
\caption{Globular cluster luminosity function for NGC 5128.
The solid dots show the magnitude distribution for the ``unbiased''
sample of 124 clusters described in the text.  Adding back in the
other 25 bright clusters that were the target centers of our ACS
and STIS exposures gives the open circles shown.  The solid curve
drawn through the data is not a fitted curve, but is a standard Gaussian GCLF for giant 
elliptical galaxies, with a dispersion $\sigma = 1.3$ mag and
turnover point at $T_1 = 20.35$.}
\label{fig6}
\end{figure}

\begin{figure} 
\figurenum{8} 
\plotone{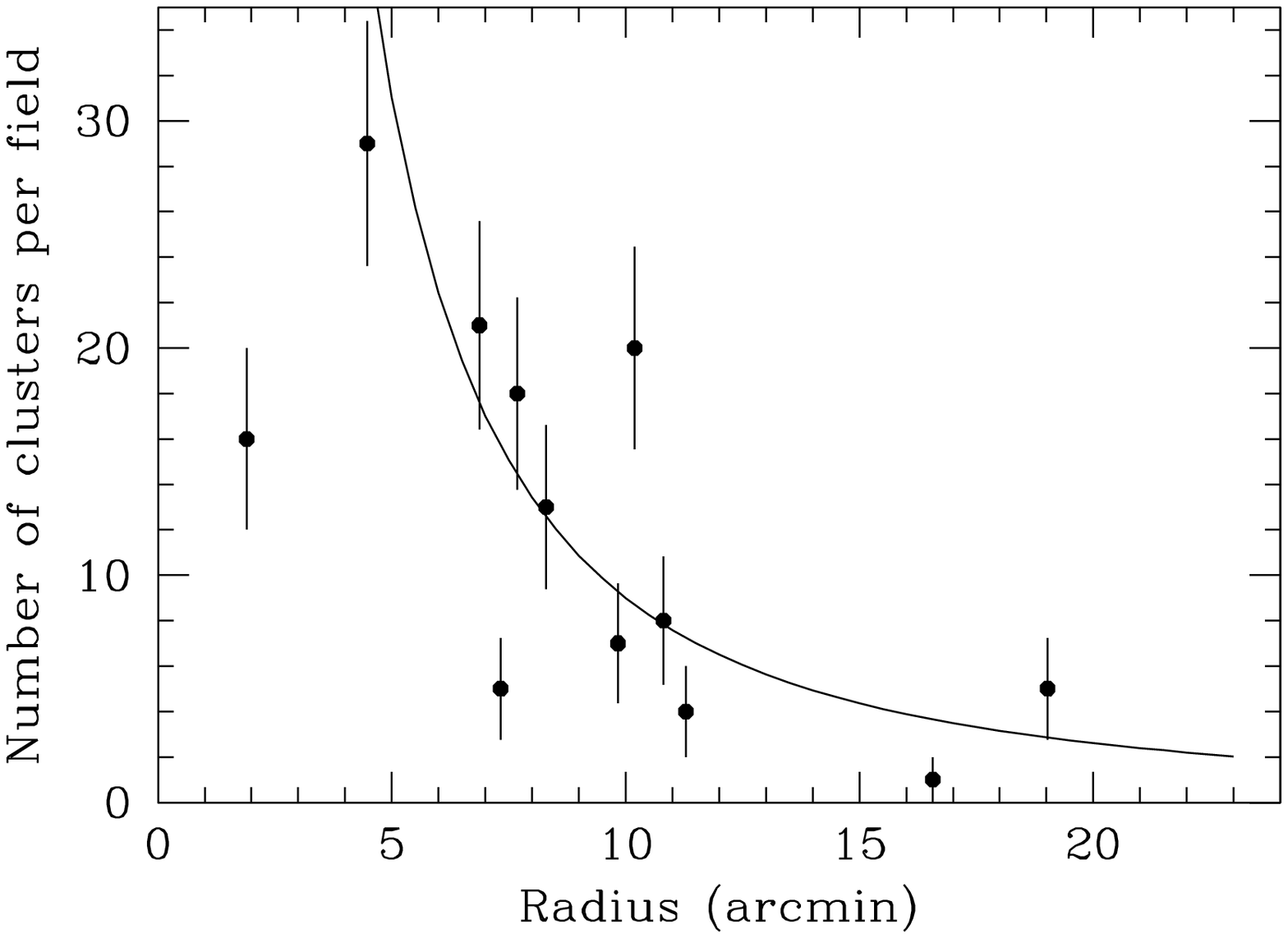} 
\caption{Radial profile for the globular cluster system in NGC 5128.
The number $n$ of detected clusters in each of our 12 ACS images is plotted
against the radius of the field from the center of NGC 5128.  Errorbars
on each point are $\pm n^{1/2}$.  The fitted
line is a power-law profile $n = 550 R^{-1.8}$ (see text).  The innermost
field, C6, is expected to be severely affected by incompleteness and crowding.}
\label{fig7}
\end{figure}

\end{document}